\documentclass[twocolumn]{article}

\usepackage{latexsym}
\usepackage{amsmath,amssymb,calc,amsfonts}
\usepackage{amsbsy}
\usepackage{mathrsfs}
\usepackage{color}
\usepackage{psfrag}
\usepackage{enumerate}
\usepackage[T1]{fontenc}
\usepackage[latin1]{inputenc}
\usepackage{hyperref}

\usepackage{graphicx,calc,epsfig}
\def\ut#1{\rlap{\lower1ex\hbox{$\sim$}}#1{}}

\newcommand{\be}{\nopagebreak[3]\begin{equation}}
\newcommand{\ee}{\end{equation}}
\newcommand{\ba}{\nopagebreak[3]\begin{eqnarray}}
\newcommand{\ea}{\end{eqnarray}}
\DeclareFontFamily{U}{rsfs}{}         
\DeclareFontShape{U}{rsfs}{m}{n}{<5> rsfs5 <6><7> rsfs7          %
  <8><9><10><10.95><12><14.4><17.28><20.74><24.88> rsfs10}{}     %
\DeclareMathAlphabet{\mathfs}{U}{rsfs}{m}{n}                     %
\newcommand{\mfs}[1]{\mathfs {#1}}                               %

\newcommand{\sE}{{\mfs E}}

\def\pb#1{\rlap{\lower1.5ex\hbox{$\longleftarrow$}}{#1}}
\def\dpb#1{\rlap{\lower1.5ex\hbox{$\Longleftarrow$}}{#1}}
\def\spb#1{\rlap{\lower1.5ex\hbox{$\leftarrow$}}{#1}}
\def\sdpb#1{\rlap{\lower1.5ex\hbox{$\Leftarrow$}}{#1}}

\definecolor{blue}{rgb}{0,0,1}
\definecolor{green}{rgb}{0,1,0}
\definecolor{red}{rgb}{1,0,0}
\definecolor{vio}{rgb}{1,0,1}
\definecolor{ama}{rgb}{1,1,0}

\begin{document}

\title{\bf Peculiar anisotropic stationary spherically symmetric solution
of Einstein equations
}

\date{\today}

\author{Emanuel Gallo and Osvaldo M. Moreschi \\
 {\it FaMAF, Universidad Nacional de C\'ordoba,} \\
 {\it Instituto de F\'isica Enrique Gaviola (IFEG), CONICET,} \\
 {\it Ciudad Universitaria, (5000) C\'ordoba, Argentina.}
}

\maketitle

\begin{abstract}
Motivated by studies on gravitational lenses, we present an exact solution of the field equations 
of general relativity, which is 
static and spherically symmetric, has no mass but has a non-vanishing spacelike components
of the stress-energy-momentum tensor.
In spite of its strange nature, this solution provides with non-trivial descriptions of gravitational effects.
We show that the main aspects found in the \emph{dark matter phenomena} can be satisfactorily described by this 
geometry.
We comment on the relevance it could have to consider non-vanishing spacelike components of the 
stress-energy-momentum tensor ascribed to dark matter.
\end{abstract}


\section{Introduction}
Although in Newtonian physics the notion of mass becomes essential for the description of gravitation,
general relativity tells us that the 
nature of gravitational phenomena is described more precisely by the geometry of the spacetime;
in which small particles follow the so called \emph{geodesic} world lines.
In the particular case of a spacetime which is spherically symmetric, it is possible to give a precise meaning
to the notion of quasi local mass. One of the most elaborated approaches was presented many year ago\cite{Penrose82},
in which the notion of mass is associated to an integral on a sphere of appropriate components of the
curvature tensor.

An important tool in the study of dark matter is the behavior of light
in the vicinity of the matter distribution.
The standard equations for the optical parameters neglect the possibility that
the spacelike components of the stress-energy-momentum tensor be non-vanishing; as for example in \cite{Schneider92}
where they suggest to use for the deflection angle the following expression
\begin{equation}\label{eq:alfa1}
 \alpha(\xi) = \frac{4G}{c^2} \int_{\mathcal{R}^2} d^2 \xi' \Sigma(\xi') \frac{\xi - \xi'}{|\xi - \xi'|^2} ;
\end{equation}
where in the thin lens approximation $\Sigma(\xi)$ is the surface mass density at position $\xi$, $G$ is the gravitational
constant and $c$ the velocity of light. 
Very recently we have deduced more general expressions\cite{Gallo11} in terms of the gauge invariant components of the curvature
tensor and the mass content $M(r)$, and found for the deflection angle of a spherically symmetric stationary spacetime
the expression
\begin{equation}\label{eq:lense-ext2-b}
\begin{split}
\alpha(J) = J \int_{-d_l}^{d_{ls}} 
 &\left[
 \frac{3 J^2}{r^2} \left( \frac{M(r)}{r^3} -  \frac{4 \pi}{3} \varrho(r) \right)
\right. \\
&\; 
\left. 
+ 4 \pi \left(\varrho(r) +  P_r(r) \frac{}{}\right)
\right] dy 
\end{split}
;
\end{equation}
where $J$ is the impact parameter of the light ray and $r=\sqrt{J^2 + y^2}$ and $y$
is a Cartesian like coordinate, in the direction of the light beam(unless for their explicit appearance,
we will use units in which $G=1$ and $c=1$.).
It is important here to observe the appearance of a term proportional to the radial component of
the stress-energy-momentum tensor; namely $P_r$, which is not taken into account in (\ref{eq:alfa1}),
since $\Sigma(\xi)$ is the projection of the mass density $\rho$ to the plane of the thin lens.
Motivated by this, we present here a peculiar solution of Einstein equations whose 
\emph{only non-zero component of the stress-energy-momentum tensor is $P_r$}.
The theoretical reasons to justify a spacetime of the nature we are presenting here may come from a variety of models,
that we will discuss below; they may include: consideration of alternatives to the cold dark matter model which study scalar or
spinor fields, also the different approaches to the problem of inhomogeneities in cosmology usually lead to
a correction to the field equations for the smooth out reference metric.

Our attitude in this article is to study a spacetime geometry that takes into account a non-zero 
spacelike component of the stress energy-momentum tensor, to see whether it could have some relevance
in astrophysical systems.

Then the key question is: does this, 
a little bit artificial spacetime, have some gravitational characteristics that
can be associated to observation?
We will show that the answer is unexpectedly affirmative, and this solution can be used to describe some properties of dark matter.

By presenting an example of a spacetime without mass content (and therefore
that it can not be associated to any kind of particle), but which it reasonably 
represents the main aspects of dark matter phenomena, we are 
pointing out that new directions might deserve attention in the study of dark matter.

\section{The geometry}
\subsection{The metric}
The geometry of a stationary spherically symmetric spacetime can be expressed in terms of the standard
line element
\begin{equation}\label{eq:ds1}
 ds^2 = a(r) \, dt^2 - b(r) \, dr^2 - r^2(d\theta^2 + \sin^2 \theta d\varphi^2) ;
\end{equation}
where it is convenient to define $M(r)$ from
\begin{equation}\label{eq:mder}
b(r) = \frac{1}{1 - \frac{2 M(r)}{r} }
;
\end{equation}
and we are using a timelike coordinate $t$, a radial coordinate $r$ and angular coordinates $(\theta,\varphi)$.

The source for this geometry, via Einstein equations, is understood in terms of an
energy-momentum tensor whose non-trivial components are
\begin{align}
 T_{tt} &=  \varrho \,  a(r) \, , \\
 T_{rr} &=    \frac{P_r}{\left( 1 - \frac{2 M(r)}{r} \right) } \, , \\
 T_{\theta \theta} &=    P_t \, r^2 \, , \\
 T_{\varphi \varphi} &=    P_t \,  r^2 \sin(\theta)^2 \, ;
\end{align}
where we have introduced the notion of radial component $P_r$ and
tangential component $P_t$, due to our general anisotropic assumption.
Here $\rho$ has the information of the mass density and $P_r$ and $P_t$ are spacelike components of the 
energy-momentum tensor.

To fix the system at this stage one normally must provide with equations of state for the matter content; that involves mathematical
relations for the stress-energy-momentum tensor components.
We choose as generalized equations of state
\begin{align}
 \varrho &= 0 \, , \\
 P_t & = 0 \, .
\end{align}

The solution of Einstein equations for this system is
\begin{align}
 a(r) &= \left(  \frac{\ln(\frac{r}{\mu})}{\ln(\frac{r_0}{\mu})}\right)^2 \, , \\
 M(r) &= 0 \, ;
\end{align}
where $\mu$ and $r_0$ are constants.
From this one can calculate the only non-vanishing component of the stress-energy-momentum tensor,
namely 
\begin{equation}
 P_r = \frac{1}{4\pi r^2 \ln(\frac{r}{\mu}) }  .
\end{equation}
At first sight one can observe that:
the geometry has a curvature logarithmic singularity at the 
internal radius $r=\mu$,
and
the metric approaches asymptotically the Minkowski value at the external 
radius $r=r_0$.
\subsection{The mass}
As we commented before, 
in Newtonian physics the \emph{notion of mass} is associated to the mechanical description of 
particles.
In general relativity instead, the notion of mass must come from the geometric properties of the
spacetime.
In particular, there is a natural notion of \emph{total mass} 
for isolated systems; represented by asymptotically flat spacetimes.
However, there is no universal notion of \emph{quasi-local mass} in general relativity;
but one of the most elaborated constructions was presented by Penrose\cite{Penrose82}
many years ago; which we have used for other purposes\cite{Moreschi04}.
Given a two-surface $S$ this construction provides the charge integrals\cite{Moreschi04}
\begin{equation}
  \label{eq:chargegeneral}
\begin{split}
  Q_S(w) = 4 \int &\left[
- \tilde w_2 (\Psi_1 - \Phi_{10}) + \right.
 2 \tilde w_1 (\Psi_2 - \Phi_{11} - \Lambda ) \\
& \left. - \tilde w_0 (\Psi_3 - \Phi_{21})
\right] dS_i^2 + {\tt c.c.}
\end{split}
\end{equation}
where, without getting into details one must only understand that quantities between 
parenthesis () are curvature components.

For a symmetric sphere $S$ in a stationary spherically symmetric spacetime, one has
$\Psi_1 = \Phi_{10} = \Psi_3 = \Phi_{21} = 0$ and
\begin{equation}
 \Psi_2 - \Phi_{11} - \Lambda = - \frac{M(r)}{r^3} {= 0 }  \, .
\end{equation}
So one can see that the geometry presented here has, strikingly, zero mass.

Since the spacetime is spherically symmetric, one has at hand simpler notions of
quasilocal mass which is specific of this geometry. From the way in which the 
quantity $M(r)$ appears in the field equations, in the standard reference frame,
one can notice that it grasps the notion of a mass. This has been observed often in
the literature, as for example in \cite{Misner:1964je}.

Adding to the properties of this spacetime one must say that although in the coordinate basis
presented, the curvature components tend to zero for large radial coordinate $r$,
the spacetime is not asymptotically flat in a technical sense\cite{Moreschi87}.
The failure not to qualify as an asymptotically flat spacetime does not come from
the curvature behavior (whose original components go to zero as $r\rightarrow\infty$) but
from the impossibility to build the conformal asymptotic metric.

\subsection{The energy conditions}
In order to see if this solution is physically acceptable one must study the so called
energy conditions. The natural question being:
Which energy conditions does this solution satisfy?
Let us recall that the {\sc weak energy condition} requires\cite{Wald84}:

 $\varrho  \geqq 0 \quad \text{,} \quad (\varrho + P_r ) \geqq 0$
and $(\varrho +  P_t ) \geqq 0$;

which {\bf is} satisfied by this solution.
The {\sc strong energy condition} requires\cite{Wald84}:

$(\varrho + P_r + 2 P_t ) \geqq 0$, $(\varrho + P_r ) \geqq 0$
and $(\varrho +  P_t ) \geqq 0$;

which {\bf is} satisfied by this solution.
While the {\sc dominant energy condition} requires\cite{Wald84}:

$\varrho  \geqq  |P_r|$
and $\varrho  \geqq |P_t|$;

which {\bf is not} satisfied by this solution.

Althought one is more comfortable with a proposed stress-energy-momentum
tensor that satisfies all known reasonable energy conditions, the point is that {two of the}  most basic energy conditions are satisfied even for
such a strange spacetime which does not have mass content.
We on purpose have
constructed a spacetime without mass, in order to show, in contrast of what it
have been considering up to know, that the only possible description of gravitational
phenomena (including dark matter) must be around the notion of mass, and
completely neglecting the spacelike components of the stress-energy-momentum tensor.
So we are presenting an extreme example of a spacetime whose stress-energy-momentum tensor has only
spacelike components and no mass content whatsoever.
Then since by design it has no mass content, it is immediate that it will not satisfy
the dominant energy condition.

The failure to satisfy the dominant energy condition normally raises fear about
maximum velocity of the matter involved; but this issue is rather complicated,
in particular a fluid model admitting tachyonic particles can still satisfy
the dominant energy condition\cite{Wong:2010qr}. 
It has also been emphasized that violation of the dominant energy condition, do not
necessarily violate causality\cite{McInnes:2001zw}.
Recently, cosmologist have speculated on the possibility of spacetimes
which violate this energy condition, in a variety of situations, as for example
in \cite{Giovannini:1999yy,Caldwell:2002ew,Carroll:2003st,GonzalezDiaz:2003rf,Ackerman:2007nb}.

In any case, we are not trying to indicate that dark matter 
would not satisfy the energy conditions, but we are trying to point out
that probably non trivial spacelike components of the stress-energy-momentum tensor
could play an important role in the description of the observations.

\section{Applying this geometry to the dark matter phenomena}

Having presented this exact solution to the field equations of general relativity 
it is natural to ask whether this solution can show some aspects of observations.
We will test this solution with three main observations that provoke the dark matter problem.
\subsection{Rotation curves of galaxies}
When studying rotation curves in galaxies, one must first remark that although this spacetime
has zero mass, the geometry is non trivial, and in particular there are circular orbits
for small particles.

Timelike geodesics must satisfy the equation
\begin{equation}\label{eq:uu}
a(r) (\frac{dt}{d\lambda})^2
- (\frac{dr}{d\lambda})^2
- r^2 (\frac{d\varphi}{d\lambda})^2 = 1
;
\end{equation}
where $\lambda$ is an affine parameter of the geodesic,
and we have already made use of the symmetry that allows
us to study just the motion in the ecuatorial plane $\theta = \frac{\pi}{2}$.
We also have the integrals of motion
\begin{equation}\label{eq:filam}
 J = r^2 \frac{d\varphi}{d\lambda},
\end{equation}
and
\begin{equation}
 E = a(r) \frac{dt}{d\lambda}=\left(  \frac{\ln(\frac{r}{\mu})}{\ln(\frac{r_0}{\mu})}\right)^2\frac{dt}{d\lambda}  .
\end{equation}
Then equation (\ref{eq:uu}) takes the form
\begin{equation}\label{eq:uu2}
  \left(  \frac{\ln(\frac{r_0}{\mu})}{\ln(\frac{r}{\mu})}\right)^2 E^2
-  (\frac{dr}{d\lambda})^2
-  \frac{J^2}{r^2} = 1 
;
\end{equation}
or
\begin{equation}\label{eq:uu3}
(\frac{dr}{d\lambda})^2
+  \left( \frac{J^2}{r^2}  -  \left(  \frac{\ln(\frac{r_0}{\mu})}{\ln(\frac{r}{\mu})}\right)^2E^2\right)  = 
-1
;
\end{equation}
from which one observes the effective potential $V_\text{ef}$
\begin{equation}
 V_\text{ef} = \frac{J^2}{2 r^2}  - \left(  \frac{\ln(\frac{r_0}{\mu})}{\ln(\frac{r}{\mu})}\right)^2\frac{E^2}{2} .
\end{equation}
The circular orbits conditions are
\begin{equation}
  \frac{J^2}{r^2}  - \frac{\ln^2(\frac{r_0}{\mu})E^2}{\ln^2\frac{r}{\mu}}  = 
- 1
,
\end{equation}
and
\begin{equation}
 0 = \frac{d^2 r}{d\lambda^2} = -\frac{d V_\text{ef}}{d r}
=
\frac{J^2}{r^3}  - \frac{\ln^2(\frac{r_0}{\mu})E^2}{r  \ln^3\frac{r}{\mu}}
;
\end{equation}
which for each $r$ constitute two conditions for the two integration
constants $J$ and $E$.
Therefore one has
\begin{equation}
J^2  = \frac{r^2 \ln^2(\frac{r_0}{\mu})E^2}{ \ln^3\frac{r}{\mu}}
,
\end{equation}
and
\begin{equation}
  E^2 \left( \frac{1}{\ln^2\frac{r}{\mu}} - \frac{1}{\ln^3\frac{r}{\mu}}    \right)
=  \ln^2(\frac{r_0}{\mu}) 
;
\end{equation}
which requires
\begin{equation}
 \ln\frac{r}{\mu} > 1 .
\end{equation}

\begin{figure}[htbp]
 \centering
\includegraphics[clip,width=0.48\textwidth]{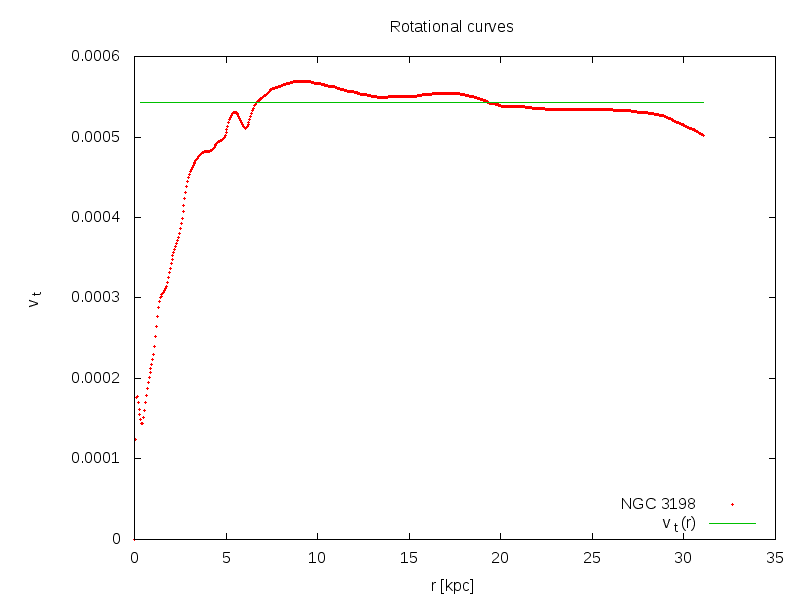}
 \caption{Observed rotation curves for NGC 3198 (red), from 
http://www.ioa.s.u-tokyo.ac.jp/\~{}sofue/RC99/3198.dat,
and calculation for a massless stress-energy-momentum tensor (green).
 }
 \label{fig:rot-curvs-3198}
 \end{figure}

\begin{figure}[htbp]
 \centering
\includegraphics[clip,width=0.48\textwidth]{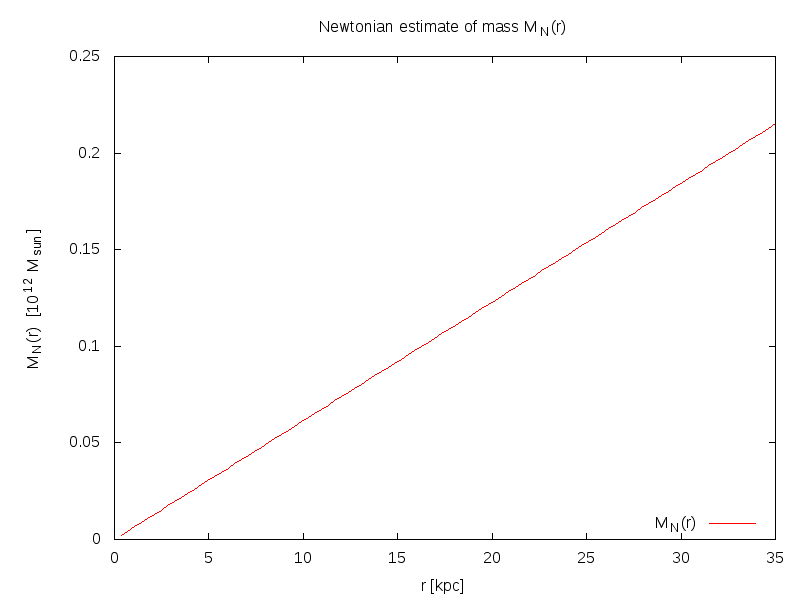}
 \caption{Newtonian estimate of the mass function $M_N(r)$, for the adapted $\mu$.
 }
 \label{fig:massN}
 \end{figure}

Let us note that in a Newtonian approach to the circular orbit problem one would deal with the equations
\begin{equation}
  \frac{J^2}{2 r^2}  - \frac{M_N(r)}{r }  = \sE
,
\end{equation}
and
\begin{equation}
 0 = \frac{d^2 r}{d t^2} = -\frac{d V_\text{ef}}{d r}
=
\frac{J^2}{r^3}  - \frac{M_N(r)}{r^2 }
;
\end{equation}
from which one would get
\begin{equation}
J^2  =  r M_N(r) 
,
\end{equation}
and
\begin{equation}
  - \frac{M_N(r)}{2 r }  = \sE
.
\end{equation}
The tangential velocity is then
\begin{equation}
 v_t(r) = r \dot \varphi = \frac{J}{r} = \sqrt{\frac{M_N(r)}{r}}
.
\end{equation}
Coming back to the original equation of motion, we see that
for circular orbits the tangential velocity $v_t$ is given by
\begin{equation}
 v_t = \frac{1}{ \sqrt{\ln(\frac{r}{\mu}) - 1 } } .
\end{equation}
It is observed that only one of the two parameters determines the rotation curve.
If one would interpret this in Newtonian terms, one would conclude that there is a mass 
content given by (putting explicitly the constants)
\begin{equation}
 M_N(r) = \frac{r \, c^2}{G} v_t^2 ;
\end{equation}
although, as we have said, the spacetime has zero mass.
Applying this to a typical galaxy with a flat rotation curve, and
adjusting the parameter to represent the rotation curve, requires $\mu$ to be
very small in the units kilo parsecs (kpc), namely $-\ln(\mu) = 3396313.01$, 
in other words one can write $\ln(\frac{r}{\mu})=\ln(\frac{r}{\text{kpc}})+3396313.01$;
and one finds the curve shown in figure \ref{fig:rot-curvs-3198}.
Using the Newtonian interpretation for this observation one would deduce a Newtonian mass function
as described in figure \ref{fig:massN}; which coincides with the linear growth predicted in the isothermal model\cite{Gallo11}.

\subsection{Gravitational lensing}
\begin{figure}[htbp]
 \centering
\includegraphics[clip,width=0.48\textwidth]{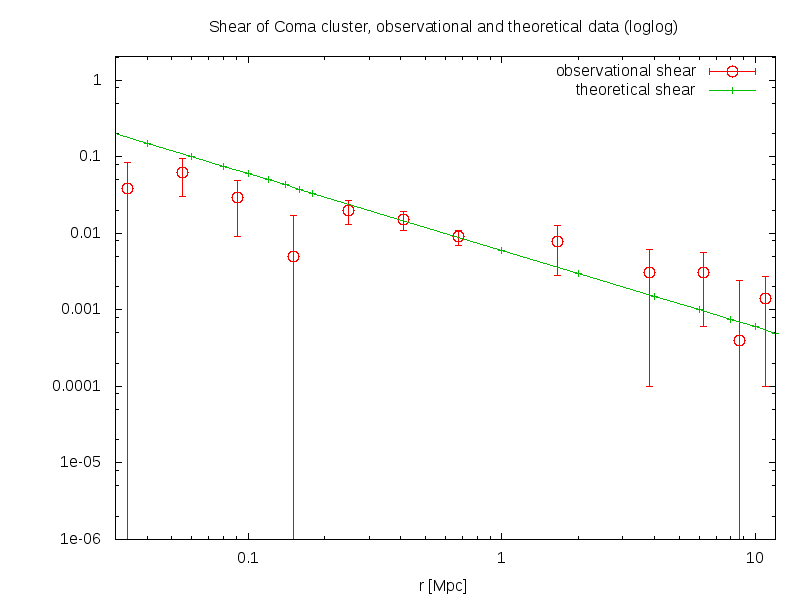}
 \caption{Observations of shear from Coma cluster for a wide range of
impact parameter as published in \cite{Kubo07} and \cite{Gavazzi:2009kf}, along with the shear calculation from
our geometry.
The first seven data points correspond to reference \cite{Gavazzi:2009kf} and the other five, for larger radii,
are from reference \cite{Kubo07}.
The observational data for the largest $r$ value from \cite{Kubo07} was excluded in this log-log graph due to the fact that it
is negative.
 }
 \label{fig:fitcomalens}
 \end{figure}

Another type of observations in which the dark matter problem is manifested is in the study
of gravitational lenses.
Let us study this geometry in the case of gravitational lensing observed
in cluster of galaxies.
Since the spacetime is not asymptotically flat one could either
match the geometry with an external metric which is asymptotically flat, for example {Minkowski metric}, 
at the external radius $r_0$;
or place source, lens and observer within the geometry\footnote{For the Coma cluster, discussed below, 
we take $r_0$ to coincide with the lens-source distance, that is 970Mpc.}.
The calculations of the optical scalars are carried out numerically from
the exact geodesic deviation equations\cite{Gallo11}; since due to the strange nature of the geometry it is not
clear whether the weak field or thin lens approximations are valid.
Fitting the free parameters in the geometry to observations\cite{Kubo07,Gavazzi:2009kf} from Coma cluster
one finds $-\ln(\mu) = 23025.8509$(in units of Mpc), which curve is shown in figure \ref{fig:fitcomalens}.
It is worthwhile to mention that the first four observational points in \ref{fig:fitcomalens} have the minimum number of galaxies
taken into account, and therefore have the least statistical weight. Then it is impressive that we can perfectly fit the other
points of the graph within the error bars, with this simple geometry without mass content.

\subsection{Scape velocity}
Other set of observations where the dark matter problem arises is in the estimation of escape velocity from
a matter distribution.
{Another technique that has been used to estimate the mass content of an
spherically symmetric region is associated to caustics in redshift
diagrams of galaxies in galaxy clusters\cite{Diaferio:1997mq}.

This techniques make use of the notion of averaged component of the escape velocity
along the line of sight at a radius of observation.
This lead us to study the notion of escape velocity in our solution.

The radial motion in the equations above is represented for the case in which
the constant of integration associated to the angular momentum vanishes, that is $J=0$.
Then the radial velocity is given by
\begin{equation}\label{eq:uu4b}
(\frac{dr}{d\lambda})^2
= \frac{\ln^2(\frac{r_0}{\mu})E^2}{ \ln^2\frac{r}{\mu}}  
- 1
.
\end{equation}
Assuming an outward radial motion for which the initial condition satisfies
\begin{equation}
\frac{\ln^2(\frac{r_0}{\mu})E^2}{ \ln^2\frac{r}{\mu}}  >
 1
;
\end{equation}
one observes that there will be a radius $r_1$ for which the radial velocity
vanishes and there will be a return in the motion and therefore the particle would not be able
to escape at all.
The scape velocity condition is to
choose $r_1$ to agree with the external radius $r_0$.
So, we set $E=1$, and therefore the escape velocity is
just
\begin{equation}
(v_{e})^2
= \frac{\ln^2\frac{r_0}{\mu}}{\ln^2\frac{r}{\mu}}  
- 1
.
\end{equation}

}

\begin{figure}[htbp]
\centering
\includegraphics[clip,width=0.48\textwidth]{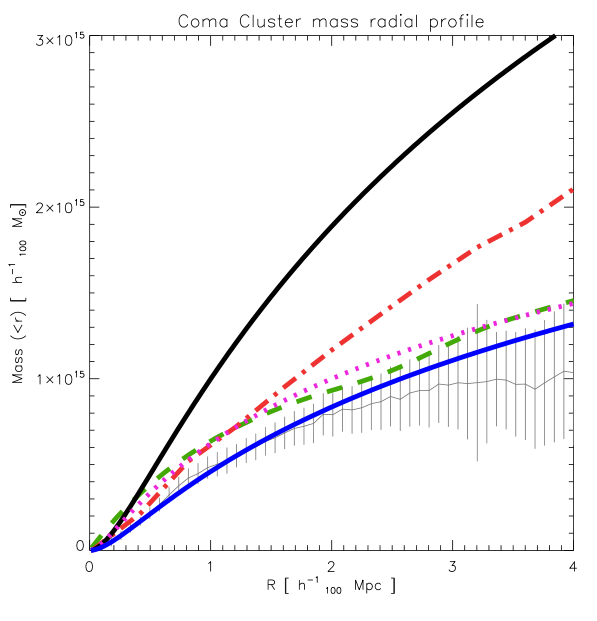}
\caption{Estimate of mass content from reference \cite{Serra:2011jh}.
For us it is only important 
the green dashed line showing the mass radial profile using caustic techniques.
}
\label{fig:serra-dominguez}
\end{figure}

\begin{figure}[htbp]
\centering
\includegraphics[clip,width=0.42\textwidth]{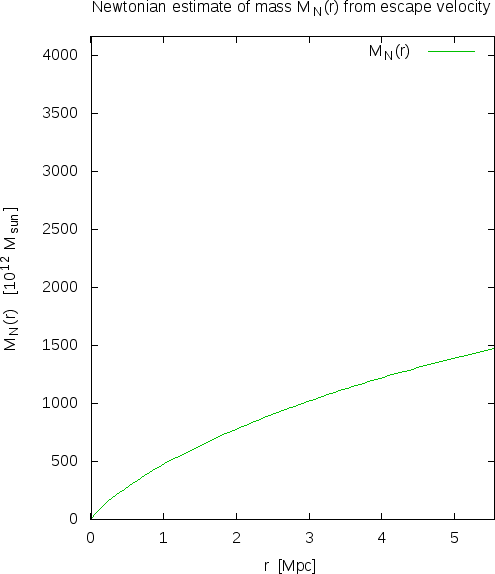}
\caption{Estimate of mass content from geometry \emph{without mass}, from escape
velocity approach; where we have chosen parameters to resemble the green dashed line of figure \ref{fig:serra-dominguez}.
}
\label{fig:escape-fit}
\end{figure}

Assuming this is due to a Newtonian distribution of mass $M_N(r)$, one would
imply a mass content of the form
\begin{equation}
 M_N(r) = \frac{r \, c^2}{2 G_N} (v_{e})^2 = 
\frac{r \, c^2}{2 G_N} \left[ \frac{\ln^2\frac{r_0}{\mu}}{\ln^2\frac{r}{\mu}} - 1\right] .
\end{equation}
In figures \ref{fig:serra-dominguez} and \ref{fig:escape-fit} one can find the estimates of mass content coming from calculations
using caustic techniques\cite{Serra:2011jh} and our fit for the same problem.
It is observed that we can reasonably represent the green dashed line estimate of reference \cite{Serra:2011jh}
with the scape velocity calculation in our geometry.

\section{Final remarks}
The physical motivation for our work comes from the need to provide a better description of the
dark matter phenomena; since in particular the estimated distribution of mass that comes
form dynamical studies and from weak lens studies do not agree.
Then noting the difference between the correct equations for the optical scalars
and the ones that have been used up to now, mentioned in the introduction,
we have constructed a spacetime that stresses this difference.

The spacetime presented here is an exact solution to Einstein equations, that 
represents an stationary spherically symmetric geometry which has \emph{zero mass}
but it satisfies the strong energy condition.
The only non-zero component of the stress-energy-momentum tensor is spacelike, with zero timelike
component, contrary to the usual assumptions.
We have shown that the main aspects of dark matter can be represented by
this peculiar geometry.
This spacetime is not intended to be the final solution of the dark matter problem;
since in particular it does not contain the contribution of visible matter.
But this geometry indicates that the problem of dark matter might need of a broader approach.

In the standard treatment of dark matter one usually assumes $(\rho \neq 0, P_r = 0, P_t = 0)$; instead we have here
studied {the opposite extreme case of a} geometry determined by $(\rho=0, P_r \neq 0, P_t = 0)$, and found that it reasonably represents basic
behavior of dark matter.
This invites us to search for the equation of state of dark matter phenomena in new directions.
For example, although there are many indications that are interpreted as pointing out to a cold dark matter model, 
if one assumes that dark matter instead of being represented by a cold non-relativistic distribution
of noninteracting particles is actually better depicted by a scalar or spinor field, then its stress-energy-momentum tensor
would contain non-trivial spacelike components.

From another point of view, one notices that,
the fact that our Universe presents an homogeneous and isotropic behavior at large distance in the past,
and a lumpy nature at short distance,
poses the problem of how to deal with the geometry and physical processes.
Although in the standard treatment of cosmological problems it is normally assumed that the exact
solution given by the Friedmann-Robertson-Walker line element is the appropriate geometry that can be applied as 
a background metric;
several ideas have been studied in this connexion, that we would like to group in three approaches:
{\bf Averaging approach (plain):} They tackle the plain idea of averaging the geometry; but one should 
decide what to average: the metric, the connexion, the curvature, or something else, as for example the trajectory of photons%
\cite{Coley:2005ei,Coley:2006kp,Paranjape:2007wr,Wiltshire:2008sg,VanDenHoogen:2008en,Paranjape:2008jc,Ellis2009,
vandenHoogen:2009nh,Rasanen:2009uw,Gasperini:2011us,Ellis:2011hk}.
{\bf Short wave limit approach:} If there is a short wave limit component to the geometry one can define precise formal
limits in this regime\cite{Burnett:1989gp,Green11}.
{\bf Multiscale approach:} If there are at least two characteristic scales in the physics of the problem, and therefore 
in the geometry of the spacetime; one should have a formalism to treat this multiscale 
situation\cite{Brill:1964zz,Isaacson:1968zz,Isaacson:1968zza,Ellis:2005uz,Wiltshire:2007fg,Wiegand:2010uh,Buchert:2011yu}. 
Some of these articles show an overlap of the approaches.
In all of these formulations there is a smoothest metric that does not satisfy Einstein equation but a 
compensated version of it; which normally includes spacelike components of the {effective} stress-energy-momentum tensor,
as we have used in the geometry presented here.

{
If one considers these type of approaches, in which there is an effective very large scale metric,
then the issue of the energy conditions that the corresponding effective very large scale stress-energy-momentum
tensor should satisfy, changes completely.
In this work we have indicated that an effective
negligible (zero) density in comparison with the effective spacelike components of the stress-energy-momentum
tensor might deserve consideration.} 


Summarising,
the geometry presented here
is not intended to be the complete description of the dark matter problem,
but indicates that tiny contributions to the geometry determined by the barionic
mass distribution might provide reasonable description of the dark matter phenomena.
We have here study the extreme situation in which the barionic mass contribution has been
completely neglected. The more physical geometry that have contribution of barionic mass
is under study, and will be presented elsewhere.
We are also studying the theoretical framework that could explain this tiny contributions,
coming from the detail analysis of the averaging problem in general relativity.

If the geometry explaining the dark matter phenomena has a different nature from the
standard cold dark matter paradigm, one would be force to calculate again all the
related consequences; as for example is the problem of 
the evolution of structure in the universe. This is the subject of future work.


\subsubsection*{Acknowledgements}
We are very grateful to Raphael Gavazzi for sending us the table with the observed 
shear for Coma cluster,
and  to A. L. Serra and M. J. L. Domínguez Romero
for allowing us to use their graph in this article.
We also thanks Domínguez Romero for suggesting improvements to the manuscript.
We acknowledge financial support from CONICET, SeCyT-UNC and Foncyt.

%

\end{document}